\documentstyle[aps,preprint,tighten,floats]{revtex}
\input{psfig.tex}

\begin{document}
\draft
\preprint{CU-TP-749 and CAL-605}

\title{Determining Cosmological Parameters from the Microwave Background}

\author{Arthur Kosowsky}
\address{Harvard-Smithsonian Center for Astrophysics,
60 Garden Street, Cambridge, Massachusetts~~02138}
\author{Marc Kamionkowski}
\address{Department of Physics, Columbia University,
New York, New York~~10027}
\author{Gerard Jungman}
\address{Department of Physics, Syracuse University,
Syracuse, New York~~13244}
\author{David N.~Spergel}
\address{Department of Astrophysical Sciences, Princeton University,
Princeton, New Jersey~~08544
\\and\\
Department of Astronomy, University of Maryland, College Park,
Maryland 20742}

\date{May 1996}
\maketitle

\begin{abstract}
Recently funded satellites will map the cosmic microwave background
radiation with unprecedented sensitivities and angular resolutions.
Assuming only primordial adiabatic scalar and tensor perturbations,
we evaluate how accurately experiments of this type will measure the
basic cosmological parameters $\Omega$ (the total density of the Universe),
$\Omega_b$ (the baryon density), $h$ (the Hubble constant), and
$\Lambda$ (the cosmological constant). The proposed experiments
are capable of measuring these 
parameters at the few-percent level. We briefly discuss
the generality of these estimates
and complications arising in actual data analysis.
\end{abstract}
\bigskip\bigskip
\centerline{To appear in Proceedings of the UCLA Dark 
Matter '96 Conference.}
\vfil\eject

The slight temperature fluctuations in the cosmic microwave
background (CMB), first detected by the COBE satellite in
1992 \cite{dmrorig}, contain a wealth of information about
the early Universe \cite{bondreview,hureview}. By measuring
the power spectrum of these anisotropies, we can hope to
extract information about the gross features of the
Universe: the Hubble constant $H_0 = 100 h \rm\,km\,s^{-1}
\,Mpc^{-1}$, which gives the expansion rate; 
the total energy density $\Omega$ in units of the critical
density $\rho_c = 3 H_0^2/8\pi G$, which determines the
geometry of the Universe; the cosmological constant $\Lambda$,
the energy density of empty space; and the baryon density
$\Omega_b$ in terms of the critical density. While information
on each of these parameters may be obtained from astronomical
measurements, all are notoriously difficult to determine;
current estimates are not very precise and are dominated by
systematic errors. The microwave background promises a completely
independent method of determining
all of these parameters. This article largely
summarizes the conclusions
of previously published work \cite{jungman1,jungman2} to which
we refer the reader for details and more extensive references.

Here we estimate the precision with which these cosmological
parameters will be determined by a high-resolution, 
high-sensitivity map of the microwave sky such as that
produced by the MAP satellite of NASA \cite{map} (slated for launch
in 2000) or ESA's COBRAS/SAMBA mission \cite{cobras} (planned for 2004).
The microwave sky is a statistical realization of an underlying
cosmological theory which predicts the temperature power spectrum:
\begin{eqnarray}
C(\theta)&\equiv&\left\langle {\Delta T({\bf\hat q}_1)\over T_0}
{\Delta T({\bf\hat q}_2)\over T_0}\right\rangle\nonumber\\
&\equiv&\sum_{l=2}^\infty {2l+1\over 4\pi}C_l P_l(\cos\theta),
\end{eqnarray}
where $\Delta T({\bf\hat q})/T_0$ is the fractional temperature
fluctuation in the direction $\bf\hat q$, 
${\bf\hat q}_1\cdot{\bf\hat q}_2 = \cos\theta$, $P_l$ are
Legendre polynomials, and the brackets are an ensemble
average over all observers; the mean CMB temperature
is $T_0=2.726\pm 0.010$ K \cite{firas}. A particular CMB
measurement will give an estimate for the values of the
multipole moments $C_l$. A simple model for a full-sky
mapping experiment which treats noise in each pixel as
gaussian and neglects correlations between pixels gives
an estimated standard error in measuring each $C_l$ as \cite{knox}
\begin{equation}
\sigma_l = \left(2\over 2l+1\right)^{1/2}
\left[ C_l + w^{-1} \exp(l^2 \sigma_b^2)\right],
\label{sigma}
\end{equation}
where $\sigma_b = 7.42\times 10^{-3} (\theta_{\rm fwhm}/1^\circ)$
for a gaussian beam, and the inverse weight per solid angle
$w^{-1}\equiv(\sigma_{\rm pix}\theta_{\rm fwhm}/T_0)^2$ is a
pixel-size-independent measure of experimental noise. 
At small $l$ (large angles), the error estimate
Eq.~\ref{sigma} is dominated by the first ``cosmic variance''
term, while at large $l$ (small angles) the noise
increases exponentially due to the beam width.

A given cosmological theory will predict the $C_l$ values.
In the following analysis, we consider the broad class
of theories in which the primordial perturbations were
adiabatic with roughly a power-law spectrum. This includes
all models based on inflation and encompasses many currently
popular models such as cold dark matter, mixed dark matter,
open models, and $\Lambda$-models. Outside of this class are
isocurvature models, including defect models like cosmic
strings and textures; some comments on distinguishing 
between these two classes of models are included below.
The adiabatic models are described by the following
set of parameters: $\Omega$, $h$, $\Lambda$, and $\Omega_b h^2$,
described above; the amplitudes and power-law indices of
the initial scalar and tensor perturbation spectra,
$Q$, $r\equiv Q_T/Q_S$, $n_S$, and $n_T$, along with
another parameter $\alpha \equiv dn_S/d\ln k$ which describes
the deviation of the scalar spectrum from a perfect power law;
the effective number of light-neutrino species at decoupling,
$N_\nu$; and the total optical depth through the epoch of
reionization, $\tau$. Given a set of values $\bf s$ for
these eleven cosmological parameters, we calculate the
moments $C_l(\bf s)$ using a semi-analytic technique
\cite{husugiyama,jungman2}. The power spectrum can also
be calculated by numerically evolving the relevant Boltzmann
equations; an efficient and public code for this purpose
has been provided in Ref.~\cite{seljakcode}.

Now it is straightforward to determine the precision to
which these parameters may be measured given the measurement
error estimate Eq.~\ref{sigma}. If the Universe is described
by the underlying parameter set ${\bf s}_0$, the probability
distribution for observing a CMB power spectrum best fit by
the parameter set $\bf s$ is
\begin{equation}
P({\bf s}) \propto \exp\left[ -{1\over 2}
({\bf s} - {\bf s}_0)\cdot[\alpha]\cdot({\bf s} - {\bf s}_0)
\right]
\label{prob}
\end{equation}
where the curvature matrix $[\alpha]$ is given approximately by
\begin{equation}
\alpha_{ij}=\sum_l {1\over \sigma_l^2} \left[
{\partial C_l\over\partial s_i}
{\partial C_l\over\partial s_j}
\right]_{{\bf s}={\bf s}_0}.
\label{curvature}
\end{equation}
In statistical terminology, the matrix $[\alpha]$ is known as
the Fisher information matrix \cite{tth}. The inverse of this
matrix, the covariance matrix $[{\cal C}] = [\alpha]^{-1}$,
gives estimates for the uncertainties in measuring
the parameters: when all parameters are fit simultaneously,
the variance in $s_i$ is ${\cal C}_{ii}$.
If some of the parameters are fixed by other means, the
variances on the rest are given by inverting the appropriate
submatrix of $[\alpha]$.

In Fig.~1, we display the standard errors for the
parameters $\Omega$, $\Lambda$, $h$, and $\Omega_b h^2$
given an underlying ``standard CDM'' model defined by
the parameters $\Omega = 1$, $h=0.5$, $\Omega_b h^2 = 0.01$,
$\Lambda = 0$, 3 light neutrinos, no reionization, no tensor
perturbations,
and a flat initial power spectrum of scalar perturbations
normalized to the COBE quadrupole, 
$Q=18$ $\mu$K \cite{dmr4yr}. Displayed as a function of beam size
are the standard (``1-$\sigma$'') errors obtainable
from a full-sky mapping experiment with two different noise
levels $w^{-1}=4.2\times 10^{-15}$
and $1.3\times 10^{-17}$. 
The first weight corresponds
to the 90 GHz channel of MAP while the second corresponds
to the 143 GHz channel of COBRAS/SAMBA. The large
disparity in sensitivity is due to the difference between
HEMT and bolometer technology: bolometers attain substantially
better sensitivity but require active cooling to mK temperatures.
In these frequency channels, MAP will have a nominal angular 
resolution of $0.29^\circ$ and COBRAS/SAMBA a resolution of
$0.17^\circ$. Note these error estimates assume {\it no}
information about any of the parameters, i.e. an 11-parameter
fit to the model. Our analysis assumes
no systematic errors (e.g. in 
foreground removal, beam profile measurement,
calculation of $C_l$'s) which lead to systematic
misestimates of the cosmological parameters.

\begin{figure}[t]
\centerline{\psfig{figure=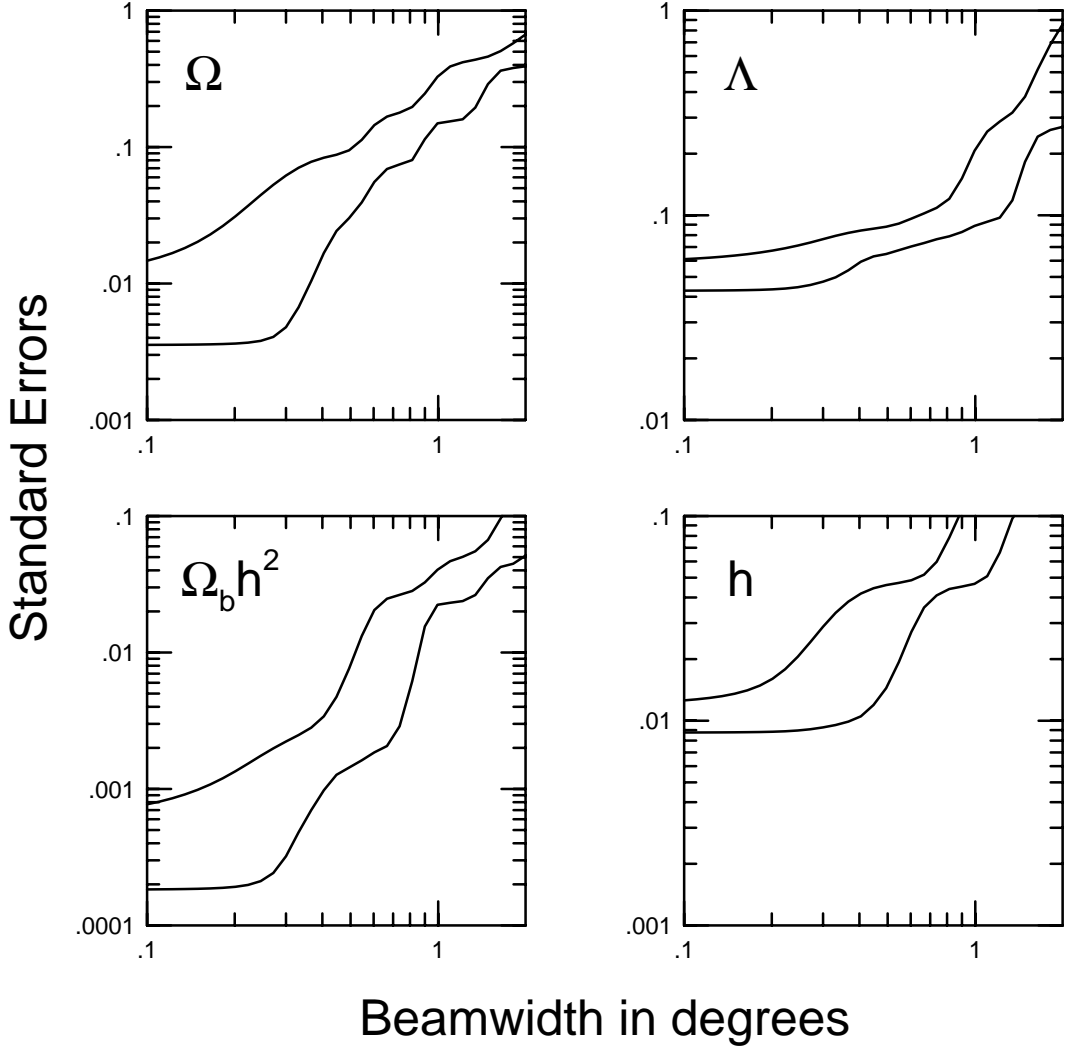,height=142mm}}
\caption{Standard errors from a
     full-sky mapping experiment as a function of beam width
     for noise levels $w^{-1}=4.2\times10^{-15}$
     (upper curve) and $1.3\times10^{-17}$ (lower curve).
     The underlying model is ``standard CDM.''}
\label{results}
\end{figure}

The CMB thus in principle offers the possibility of measuring
the basic cosmological parameters with far better precision than
traditional astronomical techniques can offer. A natural question
is the generality of these results. We have assumed a particular
cosmological model and a highly idealized experiment; what happens
when these assumptions are relaxed?

We have calculated the expected variances for
several different underlying cosmological models with
similar results, unless the universe is substantially
open or has undergone significant reionization. In the
first case, features in the power spectrum are shifted to
smaller angular scales, weakening the parameter determination
for a given beam size. However, this displacement of power
spectrum features is a robust signature of an open universe
which is difficult for any other cosmological model to mimic
\cite{kss,huwhite}, so the geometry of the universe
will still be determined to high precision. 
In the second case, if the total optical depth back to
the last scattering epoch is of order unity, features in
the power spectrum will be greatly reduced in amplitude,
hindering parameter determination. Current degree-scale
anisotropy detections from a variety of ground
and balloon experiments make this possibility unlikely
\cite{measurements,sask95}.

Structure formation may not have resulted from initial
adiabatic perturbations outside of the horizon, as with
the inflation-type models we have considered, but rather
from isocurvature fluctuations in some type of defect
model (i.e. cosmic strings or textures). We do not yet
possess highly accurate calculations of the CMB power
spectrum in such models, but recent work suggests that
the power spectrum will generically possess a substantially
different structure than in the adiabatic case 
\cite{huwhite,magueijo}. If the
Universe is actually described by cosmic strings or textures,
a microwave background map should provide an unambiguous signature,
although the extent to which cosmological parameters could 
also be determined remains an open question.

The ultimate barrier to extracting the information in the
CMB is foreground sources. At microwave frequencies,
galactic synchrotron, free-free, and dust emission are
can be of comparable amplitude to the CMB temperature fluctuations
\cite{tegmarkefst,dodelson};
in the region of the galactic plane, the foregrounds swamp
the anisotropy signal. Additionally, at angular
scales below 10' radio point sources may be a serious problem.
The galactic plane
will be cut from any CMB map as COBE did;
statistical techniques for analyzing the resulting partial-sky
map are well-known \cite{gorski}. For the rest of the
sky, the foreground signals can be removed because they
possess frequency dependences much different from blackbody.
MAP will measure in five frequency bands ranging from
22 to 90 GHz, while COBRAS/SAMBA will measure in
nine channels from 30 to 900 GHz; these frequency
spreads should be sufficient for foreground separation
at a high level of accuracy.

Finally, all of our calculations of theoretical CMB power
spectra are performed in linear perturbation theory. At
scales below a half degree, various non-linear physical
effects begin to contribute at a non-negligible level:
gravitational lensing of the microwave background by
large-scale structure \cite{seljak}; the Sunyaev-Zeldovich
effect from hot clusters \cite{sz}; and the Rees-Sciama
effect from non-linear cluster evolution \cite{rs}. While
these effects generally only give corrections to the $C_l$'s
of a few percent, the errors induced by neglecting them
are systematic. Any analysis of a high-resolution map,
particularly at angular scales below 10', should include
all of these effects for accurate parameter determination.
The estimates here are performed by truncating
the sum in Eq.~(4) at $l=1000$. The COBRAS/SAMBA
experiment can in principle make use of information
at much smaller scales than this: the lower curves
in the figure become flat below quarter-degree resolution
because the measurement becomes dominated by cosmic
variance out to $l=1000$. If theoretical 
models are understood well enough,
if foregrounds are not a serious problem, and if the
beam is determined well enough to allow probing scales
substantially smaller than the beam size, the error
estimates presented here for COBRAS/SAMBA
at the smallest angular scales could be surpassed.

In conclusion, upcoming experiments to map the cosmic
microwave background at high sensitivity and angular
resolution promise very exciting results. 
If the Universe is described by an inflation-type model with
a near power-law spectrum of initial adiabatic perturbations,
such a map will provide us with precise determination of the
basic cosmological parameters $\Omega$, $\Lambda$, $h$, and
$\Omega_b$. If on the other hand the Universe is described by
a defect model ore some other unanticipated possibility, 
these experiments will likely indicate this
unambiguously. Either way, the next decade
should bring a great increase in our knowledge of the
fundamental properties of the Universe.

\vfil\eject
\acknowledgments
This work was supported in part by the D.O.E. under contracts
DEFG02-92-ER 40699 at Columbia University
and DEFG02-85-ER 40231 at Syracuse University, by the Harvard
Society of Fellows, by the NSF under contract ASC 93-18185
(GC3 collaboration) at Princeton University,
by NASA under contract NAG5-3091 at Columbia University and
NAGW-2448 and under the MAP  Mission Concept Study Proposal at
Princeton University.

\vfil\eject

\end{document}